\documentclass{pasj00}
\draft
\def\arcmin{\ensuremath{{}^{\prime}}}

\def\farcm@mss{\ensuremath{.\mkern-4mu{}^{\prime}}}%
\def\farcs@mss{\ensuremath{.\!\!{}^{\prime\prime}}}%

\begin{document}
\SetRunningHead{K. Nariai and M. Iye}{A New Three-Mirror Anastigmat Telescope}

%\SetIndexData
%{3:{328}{},{329}{}}
%{{Kyoji}{Nariai},{Masanori}{Iye}}

\title{\LARGE Three-Mirror Anastigmat Telescope with an Unvignetted Flat Focal Plane}

%%% begin:list of authors
\author{Kyoji \textsc{Nariai}}
\affil{Department of Physics, Meisei University, Hino, Tokyo 191-8506}
\email{nariai.kyoji@gakushikai.jp}
\and
\author{Masanori \textsc{Iye}}
\affil{Optical and Infrared Astronomy Division, National Astronomical 
Observatory, Mitaka, Tokyo 181-8588}
\email{iye@optik.mtk.nao.ac.jp}

\Received{2004/9/1}
\Accepted{2004/12/17}
\Published{2005/4/25}

 %Do NOT move this preamble from here!

\maketitle

\begin{abstract}
  A new optical design concept of telescopes to provide an
  aberration-free, wide field,  unvignetted flat focal plane is
  described.  The system employs three aspheric mirrors  to remove
  aberrations, and provides a semi-circular field of view with minimum
  vignetting.  The third mirror reimages the intermediate image made
  by the first two-mirror system with a magnification factor on the
  order of unity. The present system  contrasts with the Korsch system
  where the magnification factor of the third mirror  is usually much
  larger than unity.  Two separate optical trains can be deployed to
  cover the entire circular field, if necessary.
\end{abstract}

\section{Introduction}

Most of the currently used reflecting telescopes are essentially
two-mirror systems  where major Seidel aberrations of the third order
are removed, but not entirely.   The addition of a third aspheric
mirror to construct telescope optics enables the  removal of remaining
major aberrations (\cite{paul35}; \cite{robb78}; \cite{yama83};
\cite{epps83}; \cite{schr87}; \cite{wils96}).  For instance,
\citet{will84} designed a wide-field telescope with three  aspheric
mirrors, giving $4\degree$ field of view with an image size better
than \timeform{0.''31},  or a $3\degree$ field of view with a flat
focal plane \citep{will85}. \citet{raki02a} as well as \citet{raki02b}
found solutions for a flat-field three-mirror telescope with only one
mirror  aspherized.  However, many of the three-mirror telescope
designs suffer from  obscurations, except for the design reported by
\citet{kors80} for practical applications.  A four-mirror telescope
with spherical primary is another approach to achieve small
obstruction (\cite{mein84}; \cite{wils94}; \cite{raki04}).  In the
present paper, we  report on a new concept of three-mirror telescope
design for a next-generation extremely  large telescope.

\section{Two-Mirror System}

An all-mirror optical system is characterized by the set of  the
aperture $\mathcal{D}_i$, the radius of curvature $r_i$, and  the
distance to the next surface of the $i$-th mirror surface $d_i$.  We
use $\mathcal{D}_1$ to represent the aperture diameter of the primary
mirror.

The geometry of a two-mirror system is determined  when four design
parameters ($\mathcal{D}_1$, $r_1$, $d_1$, $r_2$) are given.  We adopt
a coordinate system where light proceeds from left to right. Thus, in
a two-mirror system, $r_1<0$  and $d_1<0$; while $r_2<0$ for a
Cassegrain system  and $r_2>0$  for a Gregorian system.

Instead of radius $r$, we can also use focal length $f$.  In this
case, $f$ is positive for a concave mirror and is negative for a
convex mirror:
\begin{equation}
  f_1 = -\frac{r_1}{2},
  \qquad
  f_2 = \frac{r_2}{2}.
\end{equation}

We represent the distance from the second mirror to the focal plane of
the two-mirror system by $d_2$.  The imaging formula for the secondary
mirror gives
\begin{equation}
  \frac{1}{-(f_1 + d_1)} + \frac{1}{d_2}
  =
  \frac{1}{f_2},
\end{equation}
and the geometrical relation gives the ratio of focal lengths,
\begin{equation}
  \frac{f_1}{f_{\mathrm{comp}}}
  =
  \frac{f_1+d_1}{d_2},
\end{equation}
where $f_{\mathrm{comp}}$ is the focal length of the composite system.

For a two-mirror system,  instead of $d_2$, we often use the back
focus $d_{\rm{BF}}$, which is the distance of the focal plane from the
primary mirror,
\begin{equation}
  d_{\mathrm{BF}}=d_2+d_1 .
\end{equation}
It is more convenient to give three parameters as the focal length of
the primary mirror $f_1$,  the composite focal length
$f_{\mathrm{comp}}$, and  the focal position, which is usually
represented by $d_{\mathrm{BF}}$.  Instead of $f_1$ and
$f_{\mathrm{comp}}$, we also use  the $F$-ratio of the primary, $F_1 =
f_1/\mathcal{D}_1$, and the composite $F$-ratio, $F_{\mathrm{comp}} =
f_{\mathrm{comp}} /\mathcal{D}_1$.  When $F_1$, $F_{\mathrm{comp}}$,
$d_{\mathrm{BF}}$, and the diameter of the primary $\mathcal{D}_1$ are
given as design parameters, $f_1$, $f_{\mathrm{comp}}$, $r_1$, $d_1$,
and $r_2$ are derived by
\begin{equation}
  f_1 = F_1  \mathcal{D}_1 ,
\end{equation}
\begin{equation}
  f_{\mathrm{comp}} = f_1  \frac{F_{\mathrm{comp}}}{F_1} ,
\end{equation}
\begin{equation}
  r_1 = -2 f_1 ,
\end{equation}
%\displayadjust
\begin{equation}
  d_1 
  =
  -f_1
  \frac{f_{\mathrm{comp}} - d_{\mathrm{BF}}}{f_{\mathrm{comp}} + f_1} ,
\end{equation}
and
\begin{equation}
  r_2 
  =
  - 2 f_{\mathrm{comp}}
  \frac{f_1 + d_1} {f_{\mathrm{comp}} -f_1} .
\end{equation}

Besides the radius, a surface has parameters to characterize its
figure: the conic constant, $k$, and the coefficients of higher-order
aspheric terms.  We describe in the following the design principle of
basic two-mirror optics  systems characterized by $k_1$ and $k_2$,
since the third-order aberration coefficients are governed by these
parameters.

In a two-mirror system, because $k_1$ and $k_2$ are determined as
functions  of $r_1$, $r_2$, and $d_1$, the astigmatism $C$ and the
curvature of field $D$ are also functions of $r_1$, $r_2$,  and $d_1$.
\citet{schw05} provided an anastigmat solution with a flat field  as
$d = 2f$ and $r_1 = 2\sqrt{2}f$, where $f$ is the composite focal
length,  and is given by $1/f = 2/r_1 + 2/r_2 - 4d/r_1r_2$. Two
concentric sphere  system with $d_1 = -r_1 (1+\sqrt{5})/2$ also gives
an anastigmat solution,  known as Schwarzschild optics, which is  used
in the microscope  objective \citep{burc47}.  However, if the radii
and the distance between the mirrors are to be determined  by other
requirements, such as the final focal ratio and the back focal
distance, we can  no longer make a two-mirror system anastigmatic.

The classical Cassegrain or Gregorian telescope uses a paraboloid for
the primary ($k_1=-1$) and a hyperboloid for the secondary ($k_2 \neq
-1$).   This arrangement makes it possible to have a pinpoint image on
the optical axis by making the spherical aberration $B = 0$ by
appropriately choosing $k_2$, but the field of view is limited by
remaining non-zero coma $F$.

A Ritchey--Chr{\'e}tien telescope has no spherical aberration or coma.
The use of hyperboloids for both the primary and secondary makes it
possible to vanish two principal aberrations ($k_1$ and $k_2$ are used
to make $B=F=0$).  Most of the modern large telescopes adopt the
Ritchey--Chr{\'e}tien system  because of its wider field of view
compared with the classical Cassegrain or Gregorian  system.

The remaining aberrations among the third-order aberrations, excepting
the distortion,  are the astigmatism, $C$, and the curvature of field,
$D$.  With $k_1$ and $k_2$ already used to make $B=F=0$ for the
Ritchey--Chr{\'e}tien system, there are no  parameters available to
control $C$ and $D$.  It is thus clear that we cannot make an
anastigmat  with a two-mirror system except for some special cases
(\cite{schw05}; \cite{burc47}).

\section{Three-Mirror System}

In our three-mirror system, the last mirror is placed so that it
refocuses the intermediate image made by the first two mirrors.  The
intermediate focal plane of the first two mirrors is used as a virtual
third  surface, and the last concave mirror is numbered as the fourth
surface.

Three additional free parameters introduced are $d_3$, $r_4$, and $k_4$.
We determine $r_4$ by the condition that the Petzval sum $P$ vanishes,
\begin{equation} 
  \frac{1}{r_1} - \frac{1}{r_2} + \frac{1}{r_4} = 0 .
\end{equation} 
For ordinary telescopes, because the radius of curvature of the
primary mirror  is larger than that of the secondary mirror, $1/
|r_1|$ is smaller  than $1 / |r_2|$.  Therefore, the sign of the sum
of the first two terms is determined by $r_2$. Therefore, $r_2$ and
$r_4$ should have the same sign.  Since $r_4 < 0$, $r_2 < 0$,
the first two-mirror system should be of the Cassegrain
type, not of the Gregorian type.

The distance $d_4$ from the third mirror  to the final focal position
is calculated by
\begin{equation}
  \frac{1}{d_3} + \frac{1}{|d_4|}
  =
  \frac{2}{|r_4|} .
\end{equation}
 
The magnification factor $M$ by the third mirror is given by
\begin{equation}
  M = \left|\frac{d_4}{d_3}\right| . 
\end{equation}

Using $k_1$, $k_2$, and $k_4$, we can make $B=F=C=0$.  When we set
$P=0$, we automatically have $D=0$.  We thus obtained an anastigmatic
optical system with a flat focal plane.  As for the treatment of
higher order aberration and an  evaluation of the optical systems,
readers might wish to see some reviews (e.g., \cite{wils96};
\cite{schr87}).  We have not attempted to obtain explicit mathematical
expressions for the aberration coefficients of this three-mirror
system, but used an optimization procedure provided by the optical
design program \textit{optik} written by one of the authors (K.N.)\@.
Since the sixth-order aspheric coefficients of the mirror surface
affect  the fifth-order aberration, the sixth-order aspheric
coefficient of the primary mirror is used to control the spherical
aberration of  the fifth order of the system. For this case, too, we
used the optimizing function of \textit{optik}. The mathematical
expressions  for the fifth-order aberration coefficients can be found,
for instance, in \citet{mats72}.  Because there are 12 independent
fifth-order aberrations, it is not as straightforward as in the
third-order case to control them with available aspheric coefficients,
excepting the case of the fifth-order spherical aberration.  We
therefore treat only the fifth-order spherical aberration using the
sixth-order  aspheric coefficient of the primary mirror.

\section{Exit Pupil}

We now consider the position and the radius of the exit pupil before
discussing the vignetting problem.  Let us study the size of the
radius of the exit pupil first.  We take the primary mirror as the
pupil.

We write the lens equations for the second and third mirrors with the
sign convention for a single mirror; namely, $f_2$ is negative since
the mirror is convex and  $f_4$ is positive since the mirror is
concave.  Let $t_i$ and $t'_i$ be the distances that appear in the
imaging of pupil by the $i$-th lens.  Note that $t_2$ and $t_4$ are
positive, whereas  $t'_2$ is negative, since the image is imaginary
and $t'_4$ is positive, since the image is real.

The lens equation for the second mirror and the third mirror ($=
\mbox{surface 4}$) is written as
\begin{equation}
  \frac{1}{t_i} + \frac{1}{t'_i} 
  =
  \frac{1}{f_i} , 
  \quad (i = 2, 4).
  \label{eq:mirror23}
\end{equation}
Let us define $\xi_i$ and $\eta_i$ as
%\displayadjust
\begin{equation}
  \xi_i = \frac{f_i}{t_i } ,
  \quad
  \eta_i= \frac{t'_i}{t_i} ,
  \quad (i = 2, 4).
  \label{eq:xieta}
\end{equation}
Then, $\eta_i$ can be rewritten with $\xi_i$ or with $f_i$ and $t_i$ as
\begin{equation}
  \eta_i= \frac{\xi_i}{1-\xi_i} = \frac{f_i}{t_i-f_i} ,
  \quad (i = 2, 4).
\end{equation}
Since the distance between the first and the second mirror is usually
large compared to the focal length of the secondary mirror, $\xi_2$
and $\eta_2$ are small compared to unity, say, 0.1 in a typical case.
If the center of curvature of the third mirror is placed at the focal
position of the first two mirrors, $f_4 = d_4/2$ and $\xi_4$ and
$\eta_4$ may have a value  of around 0.3.  Using
\begin{equation}
  t_2=d_1,
  \quad
  t_4 = d_2 + d_3 - t'_2 =  d_2 + d_3 - f_2 (1+\eta_2) , 
\end{equation}
we can rewrite equation (\ref{eq:xieta}) explicitly as
\begin{equation}
  \xi_2 = \frac{f_2}{d_1},
  \quad
  \xi_4 
  = \frac{f_4}{d_2+d_3 - t'_2} 
  = \frac{f_4}{d_2+d_3 - \displaystyle\frac{d_1 f_2}{d_1-f_2}} ,
\end{equation}\smallskip\vspace*{-1pt}
\begin{equation}
  \eta_2 = \frac{f_2}{d_1-f_2},
  \quad
  \eta_4 = \frac{f_4}{d_2 + d_3- \displaystyle\frac{d_1 f_2}{d_1-f_2} -f_4}.
\end{equation}
The radius $R_{\mathrm{ep}}$ of the exit pupil is the radius of the
entrance pupil,  $\mathcal{D}_1/2$, multiplied by $t'_2/t_2$ and
$t'_4/t_4$,
\begin{eqnarray}
  R_{\mathrm{ep}}
  &=&
  \frac{\mathcal{D}_1}{2}\frac{t'_2}{t_2} \frac{t'_4}{t_4}
  = \frac{\mathcal{D}_1 \eta_2 \eta_4 }{2}
  \\[6pt]
  &=& \frac{\mathcal{D}_1}{2} \frac{f_2}{(d_1-f_2)} 
  \frac{f_4}{\left( d_2+d_3 - \displaystyle\frac{d_1 f_2}{d_1-f_2} -f_4 \right)}.
  \label{eq:Rpupil}
\end{eqnarray}
The position of the exit pupil is written as
\begin{equation}
  t'_4 = \frac{f_4}{1-\xi_4} = f_4 (1+\eta_4) .
\end{equation}

Note that the current optical system is not telecentric, since the
exit pupil is at a  finite distance.  The principal rays in the final
focal plane are not collimated, but are diverging in  proportion to
the distance from the optical axis.  This feature, however, will not
be a practical difficulty in designing the observational instrument,
unless one wants  to cover the entire field, filling 2\,m in diameter,
in a single optical train.

\section{Obstruction}

\subsection{Obstruction when $M>1$}

The detector unit at the final focal plane obstructs the ray bundle
that goes through the intermediate focal plane of the first two-mirror
system.  We solve this problem by using only the semi-circular half
field of view at the focal plane of the first two mirrors.  If the
magnification is $M=1$, the image on one half field at this virtual
plane is  reimaged to the other half side, where the detector can be
placed without essential vignetting.

If the magnification/minification factor, $M$, is not unity,  the
field of view without vignetting is narrowed because of obstruction.
It is easy to see that obstruction on the optical axis is  always
50\% regardless of the value of $M$.

Figure 1 illustrates the geometry, showing the third mirror, pupil
plane, virtual image  plane,  and final image plane, where the
position along the optical axis ($z$-axis) of each  surface  is
measured from the origin set at the apex of the third mirror.  In
figure~1, we assume that at the virtual focal plane of the two-mirror
system  (surface~3),  light passes from left to right above the axis
$(x>0)$, reflected at the third mirror  (surface~4), passing through
the exit pupil (surface~5), and imaged at the final focal plane  below
the optical axis $(x<0)$ (surface~6).

%\begin{savefloat}{fig1}
\begin{figure*}[b]
  \bigskip
  \begin{center}
    \FigureFile(140mm,71.93 mm){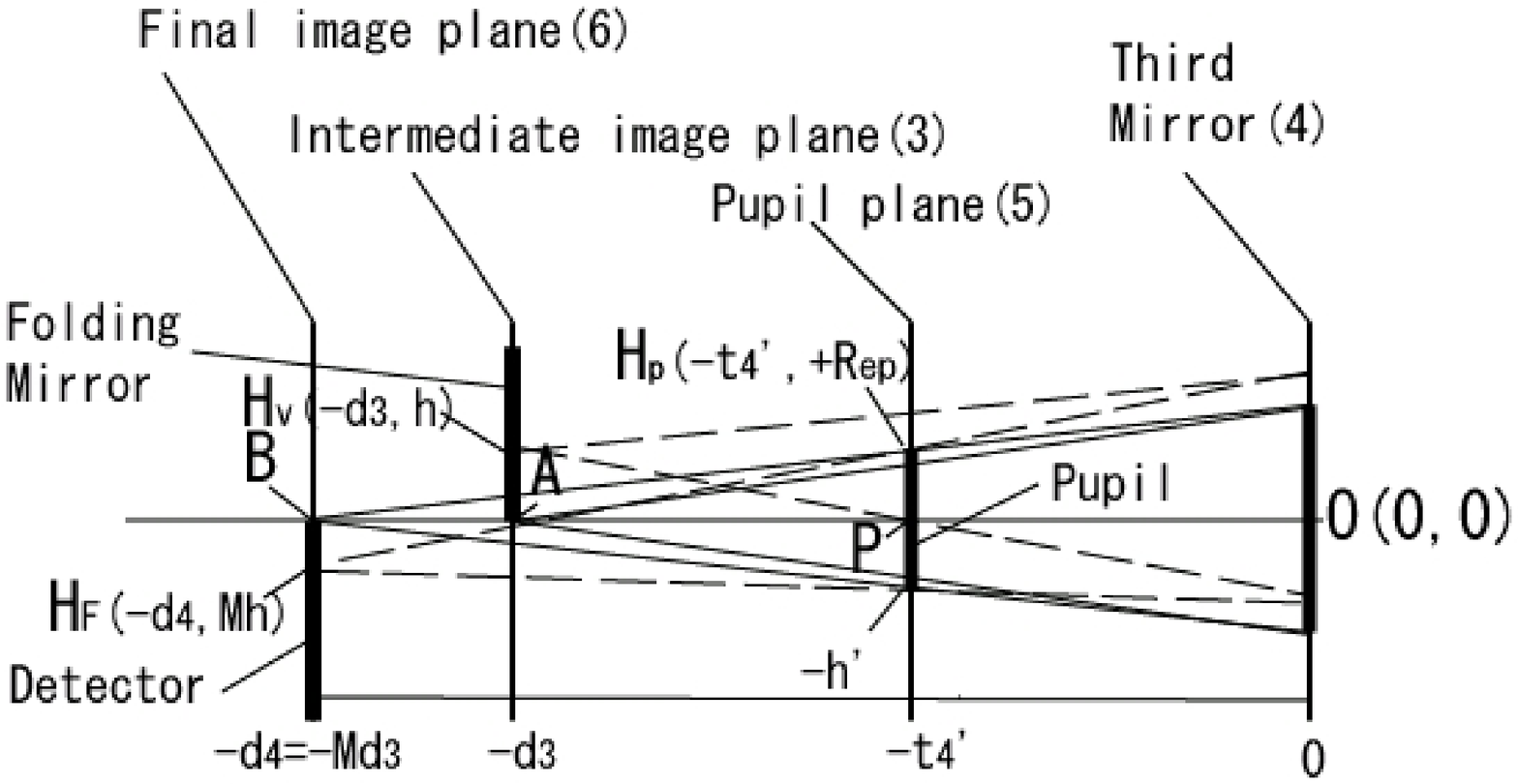} %%% 543*279
  \end{center}
  \caption{Geometry of rays defining the vignetting field for 
    $M>1$.}
  \label{pupilnw1}
\end{figure*}
%\end{savefloat}
%\putfloat{fig1}

The point A$(-d_3,0)$ at the field center of the virtual image plane
is reimaged  by the third mirror onto the point B$(-d_4,0)$ of the
final image plane.   The upper half of the beam from point A is
vignetted at the folding mirror,  and does not reach to point~B on the
detector surface.   The limiting radius, $h$, on the virtual image
plane, beyond which the beam  from the virtual plane is reflected by
the third mirror and refocused on the detector  surface,  does not
suffer any obstruction, and is defined by joining the edge point,~A,
of the folding  mirror  and the edge point
$\mathrm{H}_{\mathrm{P}}(-t'_4, R_{\mathrm{ep}})$, of the exit pupil.

Table 1 gives the coordinates of some particular points for defining
the edge of  the partially vignetted field.

%\begin{savefloat}{tab1}
\begin{table}
\caption{Coordinates of a few key points for the vignetting geometry.}
\label{poiname}
\begin{center}
\begin{tabular}{ccccl}
\hline\hline \noalign{\vskip2pt}
Point & $z$ & $x$ & Surface & Description \\[2pt]
\hline \noalign{\vskip2pt}
$\mathrm{H}_{\mathrm{V}}$ & $-d_3$ & $h$ & 3 & light enters here \\
$\mathrm{H}_{\mathrm{P}}$ & $-t'_4 $& $R_{\mathrm{ep}}$ & 5 & upper edge \\
 & & & & of the exit pupil \\  
$\mathrm{H}_{\mathrm{F}}$ & $-d_4$ & $-Mh$ & 6 & light images here \\[2pt]
\hline 
\end{tabular}
\end{center}
\end{table}
%\end{savefloat}
%\putfloat{tab1}

Because $\triangle \mathrm{BH_{F}A}$ and $\triangle \mathrm{PH_{P}A}$
are similar to each other,
\begin{equation}
  \frac{Mh}{R_{\mathrm{ep}}} = \frac{(M-1) \, d_3}{d_3-t'_4}.
\end{equation}
Thus, the limiting radius on the image plane, $Mh$, is written as
\begin{eqnarray}
  Mh 
  &=& R_{\mathrm{ep}} \frac{(M-1) \, d_3}{d_3-t'_4} \\[3pt]
  &=& R_{\mathrm{ep}}(M-1) \frac{1}{1 - \displaystyle\frac{t'_4}{d_3} } 
  =
  \frac{R_{\mathrm{ep}} (M^2-1)}{1-M\eta_4} .
\end{eqnarray}

\subsection{Obstruction when $M<1$}

In this case, the final image plane is closer to the third mirror than
is the virtual image plane, as shown in figure 2.

%\putfloat{fig2}

Because $\triangle \mathrm{BH_{V}A}$ and $\triangle \mathrm{PH_{P}B}$
are similar to each other,
\begin{equation}
  \frac{h}{R_{\mathrm{ep}}} = \frac{(1-M) \, d_3}{M d_3-t'_4}.
\end{equation}
Thus, the limiting radius on the image plane, $M h$, is written as
\begin{eqnarray}
%\displayadjust
  Mh
  &=& R_{\mathrm{ep}} M \frac{(1-M) \, d_3}{M d_3-t'_4}\\[6pt]
  &=& R_{\mathrm{ep}}(1-M) \frac{1}{1 - \displaystyle\frac{t'_4}{M d_3} } 
  = \frac{R_{\mathrm{ep}} (1-M^2)}{M-\eta_4} .
\end{eqnarray}

%\begin{savefloat}{fig2}
\begin{figure*}[b]
\bigskip
\bigskip
  \begin{center}
    \FigureFile(140mm,67.57 mm){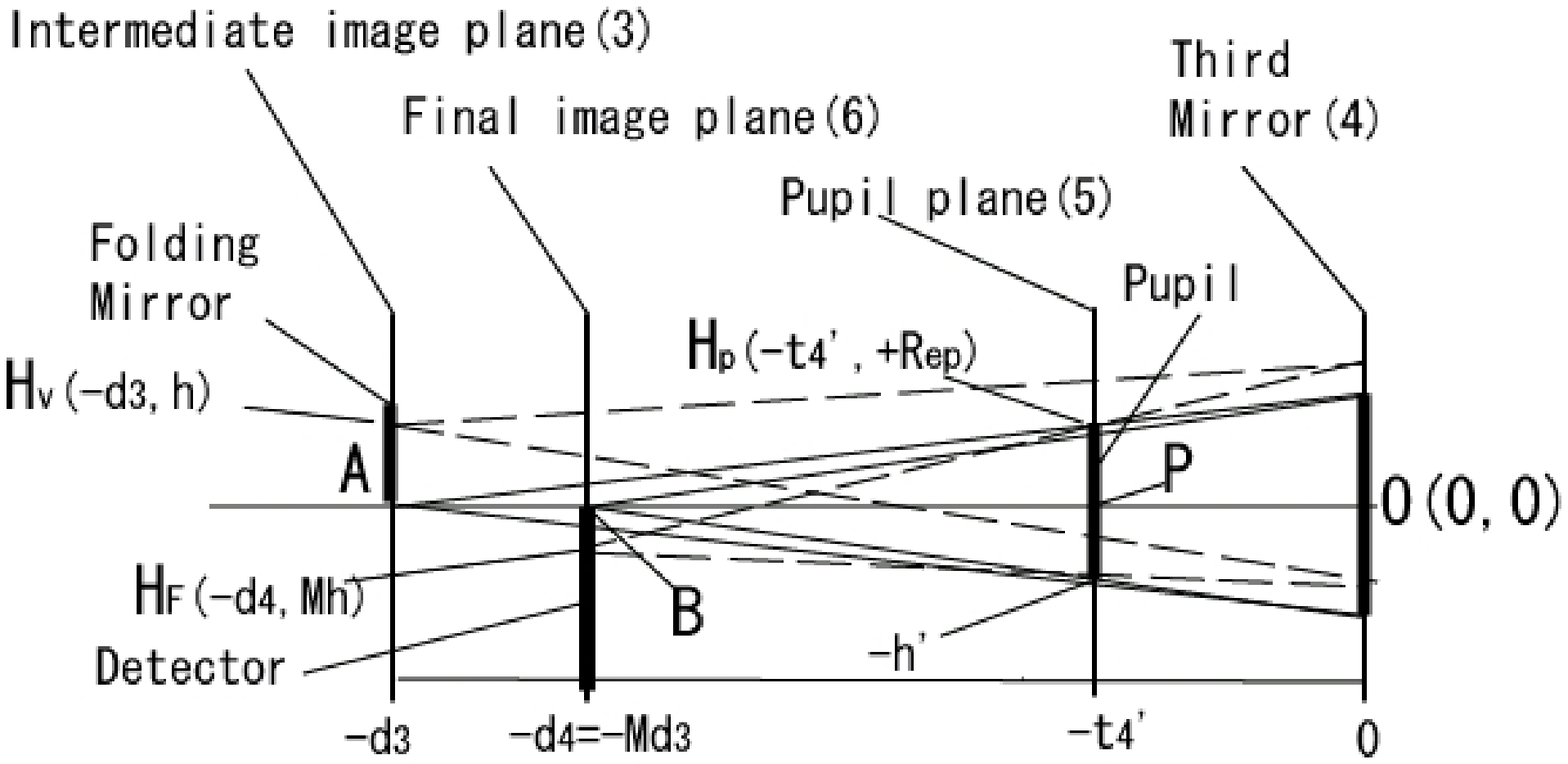} %%% 576*278
  \end{center}
  \caption{Geometry of rays defining the vignetting field for $M<1$.}
  \label{pupilnw2}
\end{figure*}
%\end{savefloat}

\subsection{Limiting Radius for the Vignetting-Free Field}

Because the 1 arcminute on the image plane is
\begin{equation}
  a
  =
  f_{\mathrm{comp}}
  \frac{\pi}{180 \times 60}
  =
  F_{\mathrm{comp}} \mathcal{D}_1
  \frac{\pi}{180 \times 60} ,
\end{equation}
the limiting radius, $Mh$, in arcminute scale, $a$, is expressed as
\begin{eqnarray}
  \frac{R_{\mathrm{ep}}  (M^2-1)}{1 - M\eta_4} \frac{1}{a} 
  = 
  \frac{1}{2F_{\mathrm{comp}}}
  \frac{M^2-1}{1-M\eta_4} 
  \eta_2 \eta_4 \frac{180 \times 60}{\pi}
  \nonumber\\[2mm]
  \hspace{25mm} (M>1) 
\end{eqnarray}
and
\begin{eqnarray}
  \frac{R_{ep}  (1-M^2)}{M-\eta_4}  \frac{1}{a} 
  =
  \frac{1}{2F_{\mathrm{comp}}}
  \frac{1-M^2}{M-\eta_4} 
  \eta_2 \eta_4 \frac{180 \times 60} {\pi} 
  \nonumber\\[2mm]
\hspace{25mm} (1>M). 
\end{eqnarray}
In a typical case, if we take the radius of field of view as
$6\arcmin$ and allow  $1\arcmin$ to be the limiting radius for the
vignetting-free field,  we have
\begin{equation}
  0.9 < M < 1.1.
\end{equation}

\begin{figure}
  \begin{center}
    \FigureFile(85mm,53.43 mm){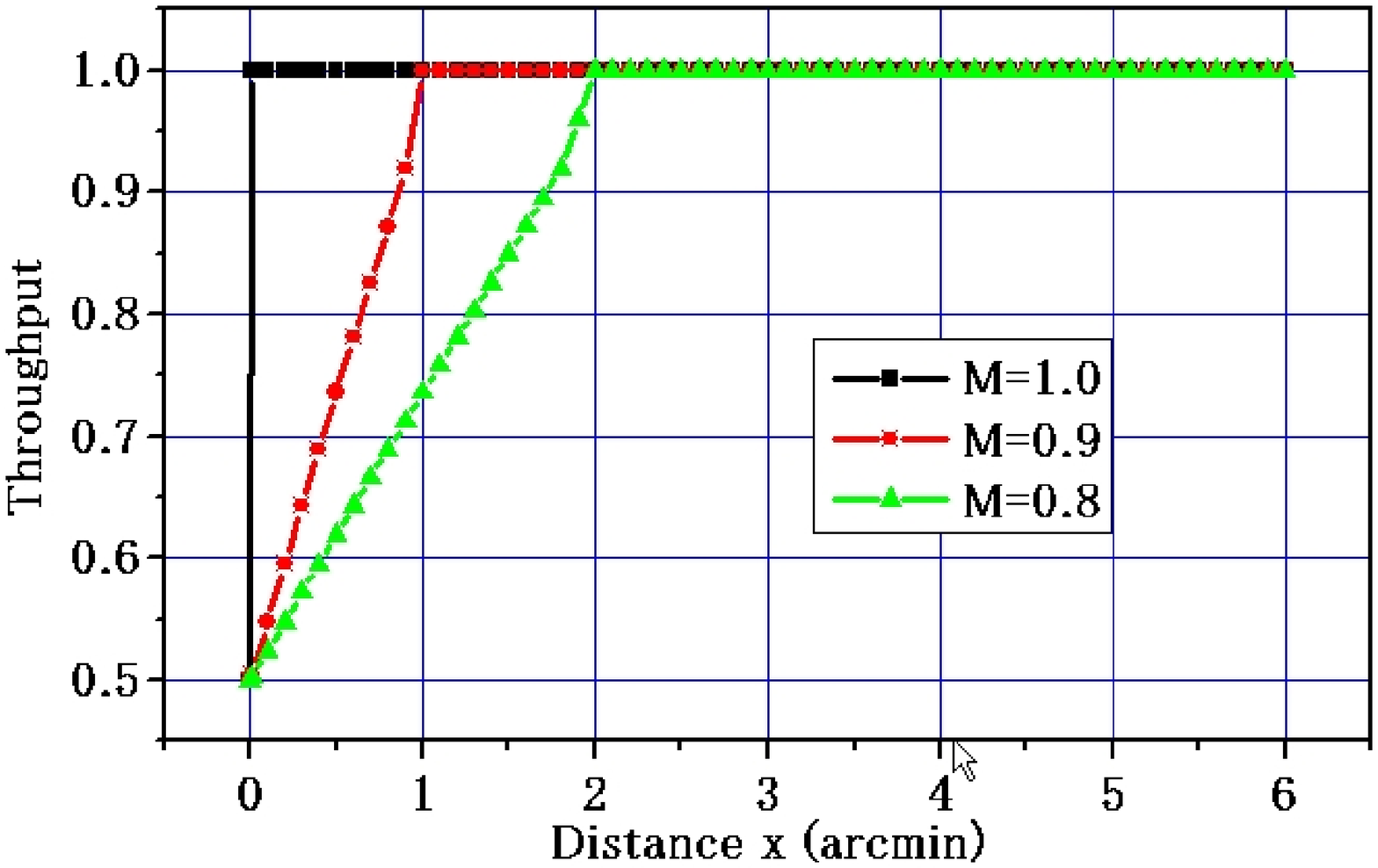} %%% 560*352
  \end{center}
  \caption{Optical throughput due to geometrical obstruction for
  $M=1$, 0.9, and 0.8.}
  \label{vignet}
  \bigskip
\end{figure}

Figure~\ref{vignet} shows the optical throughput of the present system
for three  cases with $M=1.0$, 0.9, and 0.8.  Note that for $M=1$, the
50\% vignetting takes  place only along the $x=0$ axis of the
semi-circular field. The field away from this  axis by the diffraction
size of the optics can be made essentially obstruction-free.

%\begin{savefloat}{fig3}
%\end{savefloat}
%\putfloat{fig3}

\section{Example Layout to Cover the Circular Field}

\begin{figure}[b]
  \begin{center}
    \FigureFile(85mm,65.96 mm){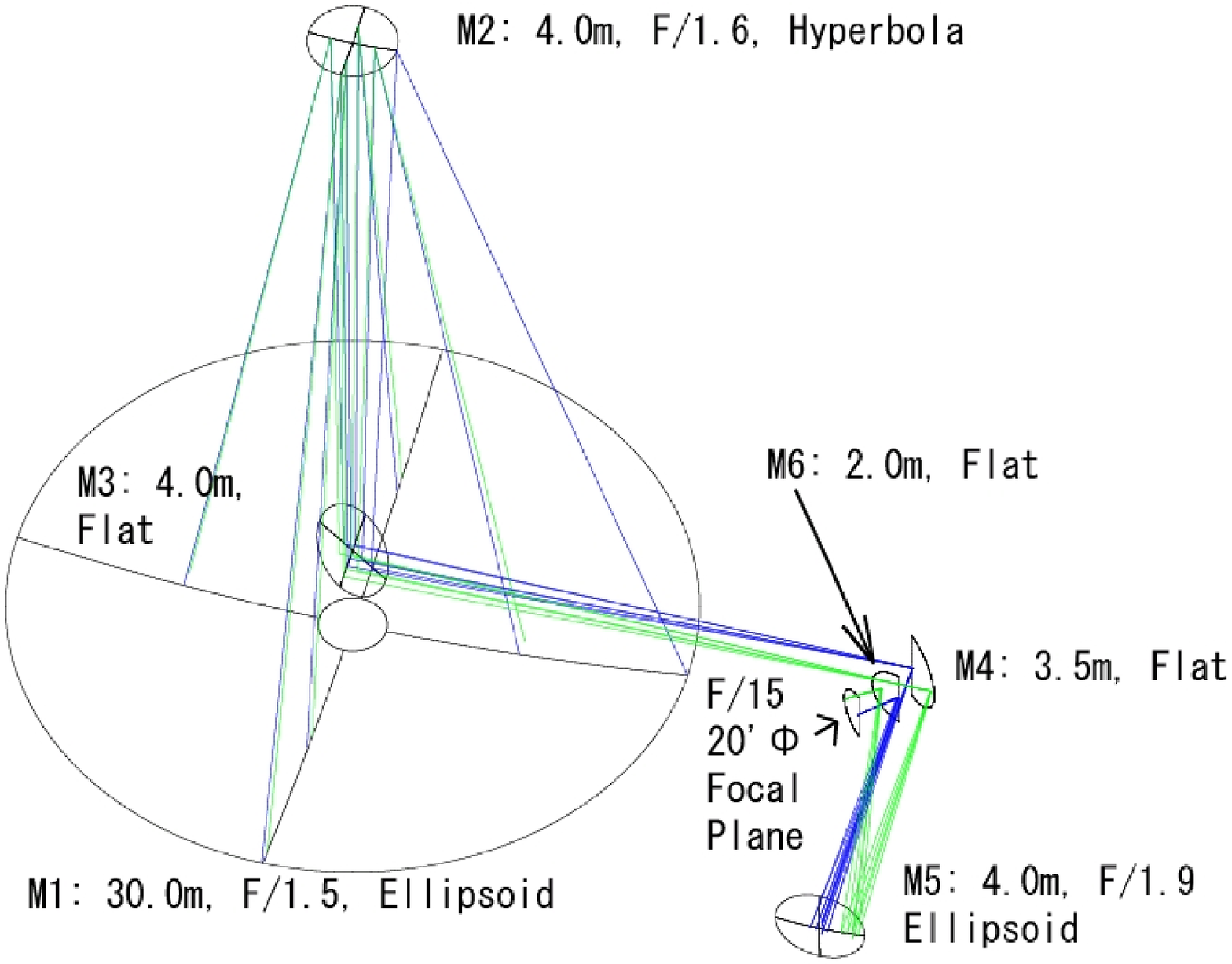} %%% 576*447
  \end{center}
  \caption{Example optical layout for a three aspheric mirror system.}
  \label{jelt3d}
\end{figure}

\begin{figure}[b]
  \begin{center}
    \FigureFile(75mm,106.56mm){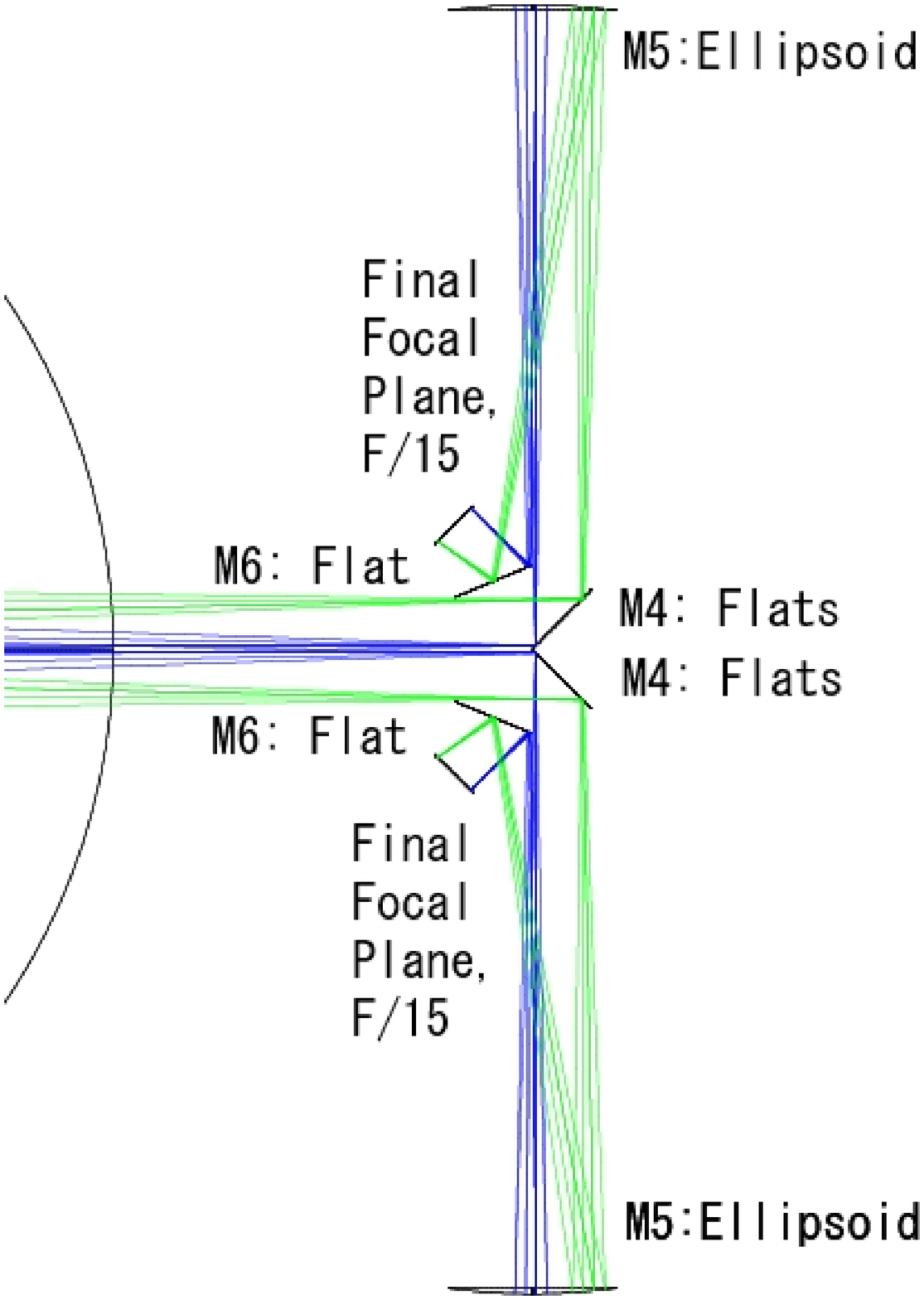} %%% 461*655
  \end{center}
  \caption{Example layout to avoid field vignetting and securing space to 
    deploy observational instruments.}
  \label{m4-6}
\end{figure}

Figure~\ref{jelt3d} shows an example layout of an all-mirror
anastigmat telescope  to  cover a full circular field of view of
$10\arcmin$  radius by two optical branches, each  covering a
semi-circular field of view.  In this figure, only one of the two
optical branches is shown, for simplicity.  By  folding the beam by
flat mirrors, M3 and M4, one can have an unvignetted semi-circular
focal plane~FP  reimaged by M5 (third aspheric mirror), and refolded
by M6 as shown in figure~\ref{m4-6}, where two optical branches are
shown.  Table~\ref{lensdat} gives the lens data of the optical system
shown in  figure~\ref{jelt3d}.

%\begin{savefloat}{fig4}
%\end{savefloat}
%\putfloat{fig4}

%\begin{savefloat}{fig5}
%\end{savefloat}
%\putfloat{fig5}

%\putfloat{tab2}

\begin{figure*}
  \begin{center}
    \FigureFile(160mm,119.88 mm){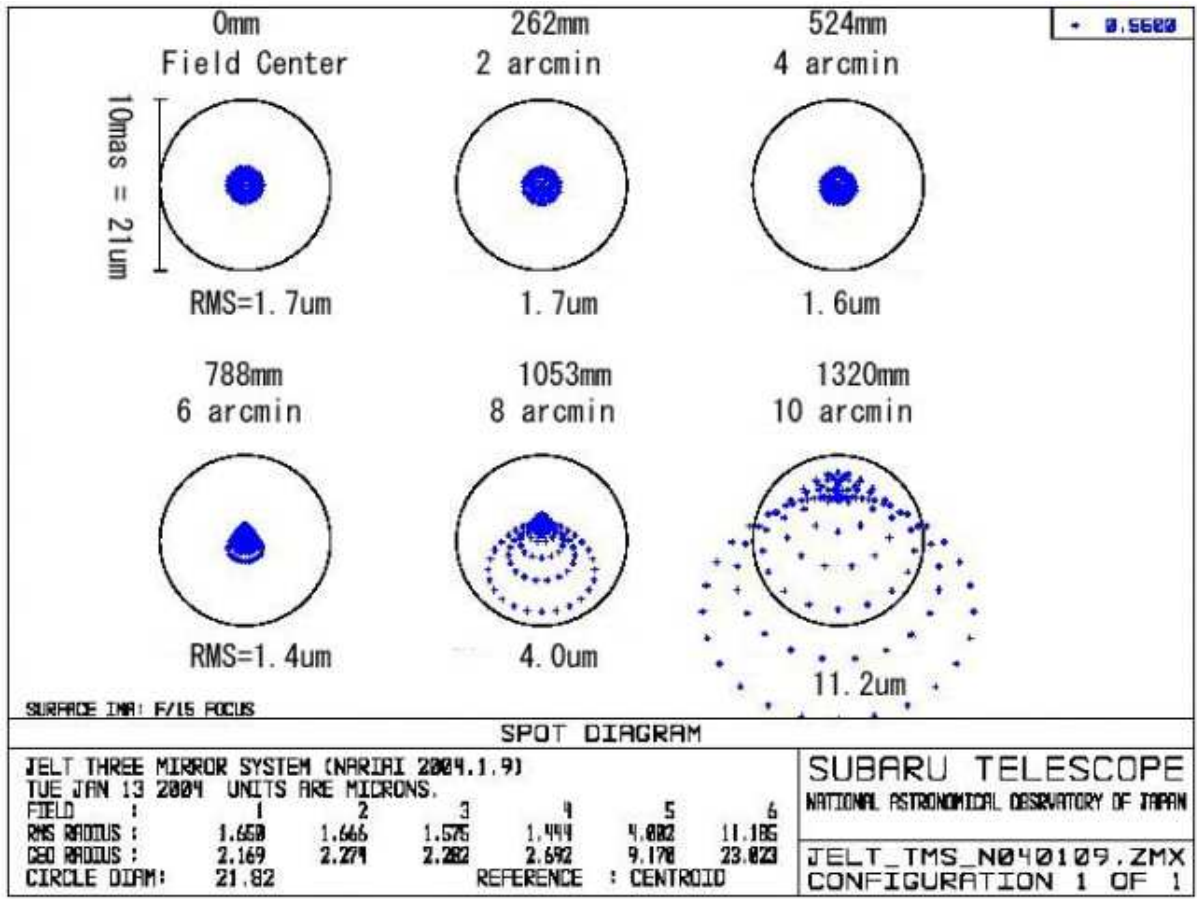} %%% 686*514
  \end{center}
  \caption{Spot diagrams up to a 10\arcmin  radius field.}
  \label{spot}
  \bigskip
\end{figure*}

A spot diagram out to $10\arcmin$ from the optical axis is shown in
figure~\ref{spot}.   Note that the designed spot sizes are smaller
than the diffraction circle for a 30\,m  ELT out to $8\arcmin$.

%\putfloat{fig6}

Actual manufacturing of such an anastigmat system needs to be further
studied.

%\breakcolumn
\section{Conclusion}

The present three-mirror anastigmat telescope system provides a flat
focal plane with  diffraction-limited imaging capability with minimal
vignetting.  The magnification factor by the third mirror, $M$,
should be designed to be close to unity.  Although only a
semi-circular field  of view can be made unvignetted in one optical
train, one can cover the entire circular field without vignetting by
deploying two such separate optical trains, each covering a
semi-circular field.

The present system is similar to the three-mirror anastigmat Korsch
system  with a $45\degree$ mirror placed at the pupil plane
\citep{kors80} concerning its aberration-free  optical performance.
However, the magnification factor by the third mirror, $M$,  should be
large in order to make the vignetting factor small for the Korsch
system,  whereas the magnification factor by the third mirror should
be close to unity in the present  system.  Therefore, the present
system can be used for applications that require a wider field  of
view.  Another merit of the present system is the avoidance of central
obscuration.

\bigskip

The authors are grateful to Dr.\ A.~Rakich, who kindly pointed out the
existence of many  important classical and modern papers to be
referred in relation to the three-mirror  telescope design.  They also
appreciate comments of Dr.\ Y.~Yamashita on the background  of the
present work.

\clearpage

%\begin{multifloat}
%\begin{savefloat}{tab2}
\begin{table*}
\caption{Preliminary lens data for the 30m JELT.}
\label{lensdat}
\begin{center} \tabcolsep4pt
\begin{tabular}{llllclllrc}
\hline\hline \noalign{\vskip2pt}
Number & Surface type & Radius & Thickness & Glass & Distance & Conic  & $\phi_v$ & \multicolumn{1}{c@{}}{$\phi_h$} & Surface\footnotemark[$\dag$] \\[2pt]
\hline \noalign{\vskip2pt}
OBJ                  & Standard               & 1.0E$+$040   & Infinity     &            & Infinity  & 0.0         &            &          & \\
1\footnotemark[$*$]  & Even Asphere           & $-90.0$      & $-39.090909$ & Mirror M1  & 15.0      & $-0.992036$ & 1.5E$-$015 & 0.0      & \\
STOP                 & Even Asphere           & $-13.131313$ & 34.090909    & Mirror M2  & 1.975613  & $-1.412689$ & 0.0        & 0.0      & \\
3                    & Coord Break            &              & 0.0          & $\cdots$   & 0.0       &             & 45.0       & 0.0      & \\
4\footnotemark[$*$]  & Standard               & 1.0E$+$040   & 0.0          & Mirror M3  & 2.112906  & 0.0         &            &          & \\
5                    & Coord Break            &              & $-25.0$      & $\cdots$   & 0.0       &             & 45.0       & 0.0      & \\
6                    & Coord Break            &              & 0.0          & $\cdots$   & 0.0       &             & 0.0        & 45.0     & \\
7\footnotemark[$*$]  & Standard               & 1.0E$+$040   & 0.0          & Mirror M4  & 1.772806  & 0.0         &            &          & 3 \\
8                    & Coord Break            &              & 15.374510    & $\cdots$   & 0.0       &             & 0.0        & 45.0     & \\
9\footnotemark[$*$]  & Even Asphere           & $-15.374510$ & $-13.374510$ & Mirror M5  & 1.999334  & $-0.720989$ & 0.0        & 0.0      & 4 \\
10                   & Coord Break            &              & 0.0          & $\cdots$   & 0.0       &             & 0.0        & $-22.5$  & \\
11\footnotemark[$*$] & Standard               & 1.0E$+$040   & 0.0          & Mirror M6  & 1.039940  & 0.0         &            &          & 6 \\
12                   & Coord Break            &              & 2.0          & $\cdots$   & 0.0       &             & 0.0        & $-22.5$  & \\
IMA                  & $-1.0\mathrm{E}{+}040$ &              &              & 1.185943   & 0.0       &             &            &          & \\[2pt]
\hline \noalign{\vskip2pt}
\multicolumn{10}{@{}l@{}}{\hbox to0pt{\parbox{175mm}{\footnotesize
$\phi_v$ denotes the angle of folding flat mirror to redirect the optical axis within the vertical plane.
\par
$\phi_h$ denotes the angle of folding flat mirror to redirect the optical axis within the horizontal plane.
\par
\footnotemark[$*$] denotes physical surfaces.
\par
\footnotemark[$\dag$] denotes the surface number corresponding to those referred in subsection 5.1 and figures 1 and 2.}\hss}}
\end{tabular}
\end{center}
\end{table*}
%\end{savefloat}

%\begin{savefloat}{fig6}
%\end{savefloat}
%\end{multifloat}


\begin{thebibliography}{}
\bibitem[Burch(1947)]{burc47}
\ADSentry{1947PPS....59...41B}{Burch, C.~R.\ 1947, Proc. Phys. Soc., 59, 41}

\bibitem[Epps(1983)]{epps83}
\ADSentry{1983AnTok..19..401E}{Epps, H.~W., \& Takeda, M.\ 1983, Ann. Tokyo Astron. Obs., 19, 401}

\bibitem[Korsch(1980)]{kors80}
\ADSentry{1980ApOpt..19.3640K}{Korsch, D.\ 1980, Appl. Opt. 19, 3640}

\bibitem[Matsui(1972)]{mats72}
Matsui, Y.\ 1972, Lens Sekkeiho (Tokyo: Kyoritu Publishing Co.),\break in Japanese

\bibitem[Meinel et~al.(1984)]{mein84}
\ADSentry{1984ApOpt..23.3020M}{Meinel, A.~B., Meinel, M.~P., Su, D.-Q., \& Wang, Y.-N.\ 1984, Appl. Opt., 23, 3020}

\bibitem[Paul(1935)]{paul35}
Paul, M.\ 1935, Rev. Opt., A14, 169

\bibitem[Rakich(2002)]{raki02b}
\ADSentry{2002SPIE.4768...32R}{Rakich, A.\ 2002, Proc. SPIE, 4768, 32}

\bibitem[Rakich and Rumsey(2002)]{raki02a} 
\ADSentry{2002OSAJ...19.1398R}{Rakich, A., \& Rumsey, N.\ 2002,  J. Opt. Soc. Am. A, 19, 1398}

\bibitem[Rakich, Rumsey(2004)]{raki04}
\ADSentry{2004SPIE.5249..103R}{Rakich, A., \& Rumsey, N.~J.\ 2004,  Proc. SPIE, 5249, 103}

\bibitem[Robb(1978)]{robb78}
\ADSentry{1978ApOpt..17.2677R}{Robb, P.~N.\ 1978, Appl. Opt., 17, 2677}

\bibitem[Schroeder(1987)]{schr87}
\ADSentry{1987asop.book.....S}{Schroeder, D.~J.\ 1987, Astronomical Optics (New York: Academic Press)}

\bibitem[Schwarzschild(1905)]{schw05}
Schwarzschild, K.\ 1905, Investigations into Geometrical Optics II, Theory of Mirror Telescopes, English translation by A.\ Rakich, $\langle$http://members.iinet.net.au/\textasciitilde{}arakich/$\rangle$

\bibitem[Willstrop(1984)]{will84}
\ADSentry{1984MNRAS.210..597W}{Willstrop, R.~V.\ 1984, MNRAS, 210, 597}

\bibitem[Willstrop(1985)]{will85}
\ADSentry{1985MNRAS.216..411W}{Willstrop, R.~V.\ 1985, MNRAS, 216, 411}

\bibitem[Wilson(1996)]{wils96}
\ADSentry{1996rtob.book.....W}{Wilson, R.~N.\ 1996, Reflecting Telescope Optics I, Basic Design Theory and its Historical Development (Berlin: Springer-Verlag)}

\bibitem[Wilson et~al.(1994)]{wils94}
\ADSentry{1994SPIE.2199.1052W}{Wilson, R.~N., Delabre, B., \& Franza, F.\ 1994, Proc. SPIE, 2199, 1052}

\bibitem[Yamashita, Nariai(1983)]{yama83}
\ADSentry{1983AnTok..19..375Y}{Yamashita, Y., \& Nariai, K.\ 1983, Ann. Tokyo Astron. Obs., 19, 375}
\end{thebibliography}
\end{document}